\documentclass[12pt]{iopart}

\usepackage{graphicx}
\usepackage{dcolumn}
\usepackage{caption}

\begin{document}
\title[]{Two-particle angular correlations in p+p and Cu+Cu collisions at PHOBOS}
\author{Wei Li for the PHOBOS collaboration}

\address{
\vspace{2mm}
{\scriptsize
Massachusetts Institute of Technology, 77 Mass Ave, Cambridge, MA 02139-4307, USA\\
}}
\ead{davidlw@mit.edu}
{\footnotesize
B.Alver$^4$,
B.B.Back$^1$,
M.D.Baker$^2$,
M.Ballintijn$^4$,
D.S.Barton$^2$,
R.R.Betts$^6$,
A.A.Bickley$^7$,
R.Bindel$^7$,
W.Busza$^4$,
A.Carroll$^2$,
Z.Chai$^2$,
V.Chetluru$^6$,
M.P.Decowski$^4$,
E.Garc\'{\i}a$^6$,
N.George$^2$,
T.Gburek$^3$,
K.Gulbrandsen$^4$,
C.Halliwell$^6$,
J.Hamblen$^8$,
I.Harnarine$^6$,
M.Hauer$^2$,
C.Henderson$^4$,
D.J.Hofman$^6$,
R.S.Hollis$^6$,
R.Ho\l y\'{n}ski$^3$,
B.Holzman$^2$,
A.Iordanova$^6$,
E.Johnson$^8$,
J.L.Kane$^4$,
N.Khan$^8$,
P.Kulinich$^4$,
C.M.Kuo$^5$,
W.Li$^4$,
W.T.Lin$^5$,
C.Loizides$^4$,
S.Manly$^8$,
A.C.Mignerey$^7$,
R.Nouicer$^2$,
A.Olszewski$^3$,
R.Pak$^2$,
C.Reed$^4$,
E.Richardson$^7$,
C.Roland$^4$,
G.Roland$^4$,
J.Sagerer$^6$,
H.Seals$^2$,
I.Sedykh$^2$,
C.E.Smith$^6$,
M.A.Stankiewicz$^2$,
P.Steinberg$^2$,
G.S.F.Stephans$^4$,
A.Sukhanov$^2$,
A.Szostak$^2$,
M.B.Tonjes$^7$,
A.Trzupek$^3$,
C.Vale$^4$,
G.J.van~Nieuwenhuizen$^4$,
S.S.Vaurynovich$^4$,
R.Verdier$^4$,
G.I.Veres$^4$,
P.Walters$^8$,
E.Wenger$^4$,
D.Willhelm$^7$,
F.L.H.Wolfs$^8$,
B.Wosiek$^3$,
K.Wo\'{z}niak$^3$,
S.Wyngaardt$^2$,
B.Wys\l ouch$^4$\\
}
\vspace{3mm}
\small

{\scriptsize
\noindent
$^1$~Argonne National Laboratory, Argonne, IL 60439-4843, USA\\
$^2$~Brookhaven National Laboratory, Upton, NY 11973-5000, USA\\
$^3$~Institute of Nuclear Physics PAN, Krak\'{o}w, Poland\\
$^4$~Massachusetts Institute of Technology, Cambridge, MA 02139-4307, USA\\
$^5$~National Central University, Chung-Li, Taiwan\\
$^6$~University of Illinois at Chicago, Chicago, IL 60607-7059, USA\\
$^7$~University of Maryland, College Park, MD 20742, USA\\
$^8$~University of Rochester, Rochester, NY 14627, USA\\
}

\begin{abstract}
We present results on two-particle angular correlations in p+p and Cu+Cu collisions over 
a broad range of ($\eta$,$\phi$). The PHOBOS detector has a uniquely large angular coverage
for inclusive charged particles. This allows for the study of correlations on both long 
and short range pseudorapidity scales. A complex two-dimensional correlation structure 
emerges which is interpreted in the context of a cluster model. The cluster size and its 
decay width are extracted from the two-particle pseudorapidity correlation function. 
Relative to p+p collisions, Cu+Cu reactions show a non-trivial decrease of cluster size 
with increasing centrality. These results may provide insight into the hadronization stage 
of the hot and dense medium created in heavy ion collisions. 
\end{abstract}


\section{Introduction}
Multiparticle correlation analyses have proven to be a powerful tool in exploring the 
underlying mechanism of particle productions in high energy hadronic collisions. Both short- 
and long-range correlations have been discovered in the past decades \cite{UA5_3energy,ISR_twolowenergy} 
which have been given a simple interpretation via the concept of clustering \cite{cluster_model,cluster_fit}.
In a cluster model, clusters of hadrons are formed first and emitted independently according 
to some basic scheme. They then decay isotropically in their center of mass frame into the 
observed hadrons. Two-particle angular correlations provide detailed information about the 
cluster properties, e.g. their multiplicity (``size'') and extent in phase space (``width''). 
In heavy ion collisions at RHIC, the expected formation of a Quark Gluon Plasma (QGP) could 
lead to a modification of the clusters relative to p+p collisions \cite{AAcluster_prediction}. 
This study should provide a useful baseline measurement for understanding hadronization stage 
in A+A collisions.

\section{Analysis method}

Covering pseudorapidity range ($\eta=-\ln(\tan(\theta/2))$) $-3<\eta<3$ over almost full 
azimuthal angle, the PHOBOS Octagon detector \cite{phobos_detector} is ideally suited for 
direct study of the angular correlations of the particles emitted from clusters. 
Following a similar approach as in Ref. \cite{ISR_twolowenergy}, the inclusive two-particle 
correlation function in ($\Delta \eta,\Delta \phi$) space is defined as follows:

\begin{equation}
\label{2pcorr_incl}
R(\Delta \eta,\Delta \phi)=\left<(n-1)\left(\frac{F_{n}
(\Delta \eta,\Delta \phi)}{B_{n}(\Delta \eta,\Delta \phi)}-1\right)\right>
\end{equation} 

\noindent where $F_{n}(\Delta \eta,\Delta \phi)$ is the foreground distribution determined 
by taking two-particle pairs from the same event and $B_{n}(\Delta \eta,\Delta \phi)$ is the 
background distribution constructed by randomly selecting single particles from two different 
events with similar vertex position and centrality. The track multiplicity, $n$, is introduced 
to compensate for the trivial dilution effects from uncorrelated particles. $R(\Delta \eta,\Delta \phi)$ 
is defined such that if a heavy ion collision is simply a superposition of individual p+p 
collisions, and thus has the same local correlations, the same correlation function should be observed.

\section{Two-particle correlations in p+p collisions}

\vspace{-0.5cm}
\begin{figure}[t]
\captionsetup[figure]{margin=0.1cm,font=small}

\begin{minipage}[t]{0.45\textwidth}
 \centerline{
  \includegraphics[width=\textwidth]{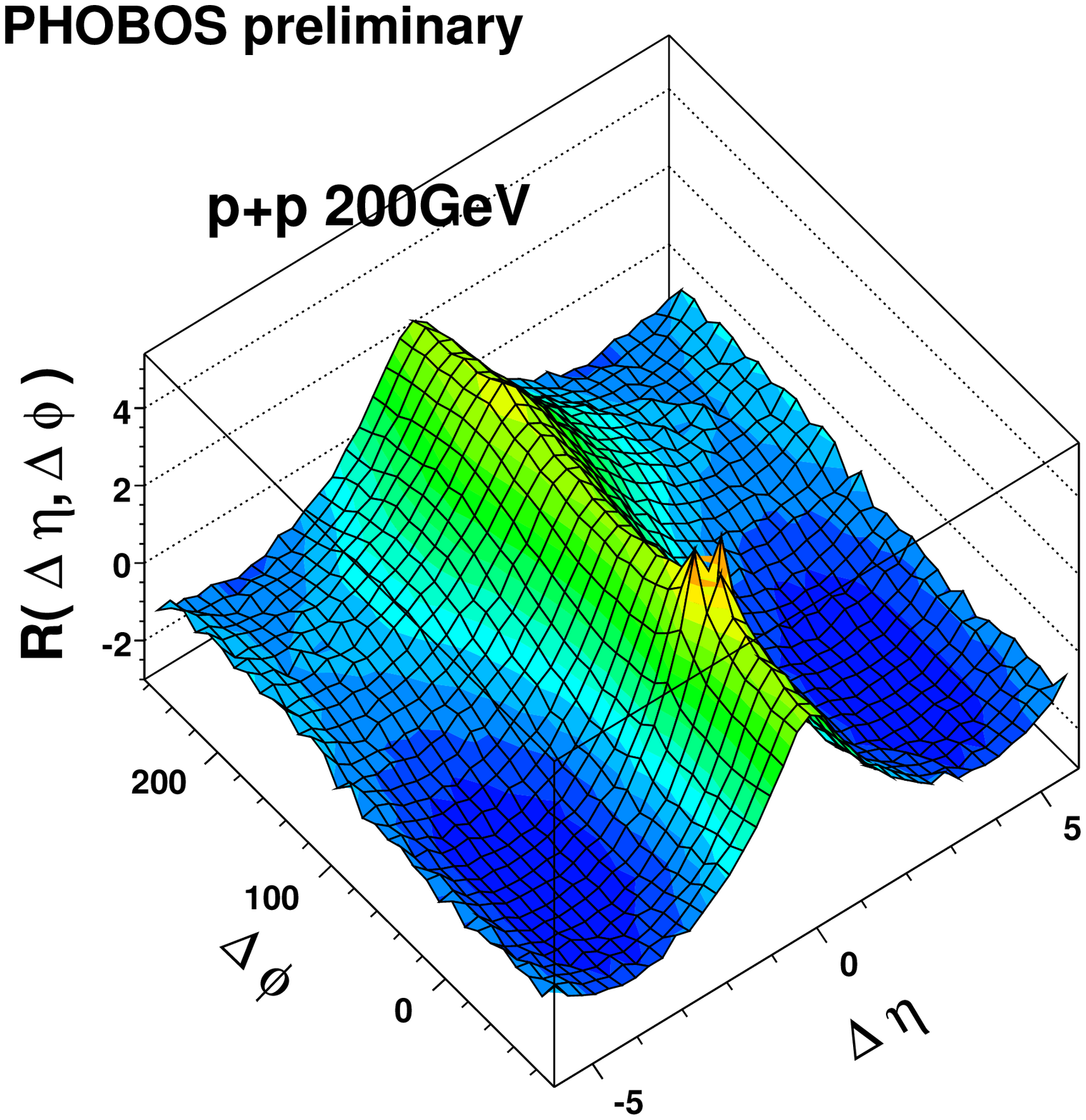}
 }
 \vspace{-0.3cm}
 \caption{Final fully corrected two-particle correlation 
          function in ($\Delta \eta$, $\Delta \phi$) for 
          p+p collisions at $\sqrt{s}$ = 200~GeV. A small
          region near ($\Delta \eta$, $\Delta \phi$)
          of (0,0) has been removed due to its large experimental uncertainty
          arising from the secondary effects.}
 \label{pp200_2D_corrected}
\end{minipage}
\hspace{\fill}
\begin{minipage}[t]{0.45\textwidth}
 \centerline{
  \includegraphics[width=\textwidth]{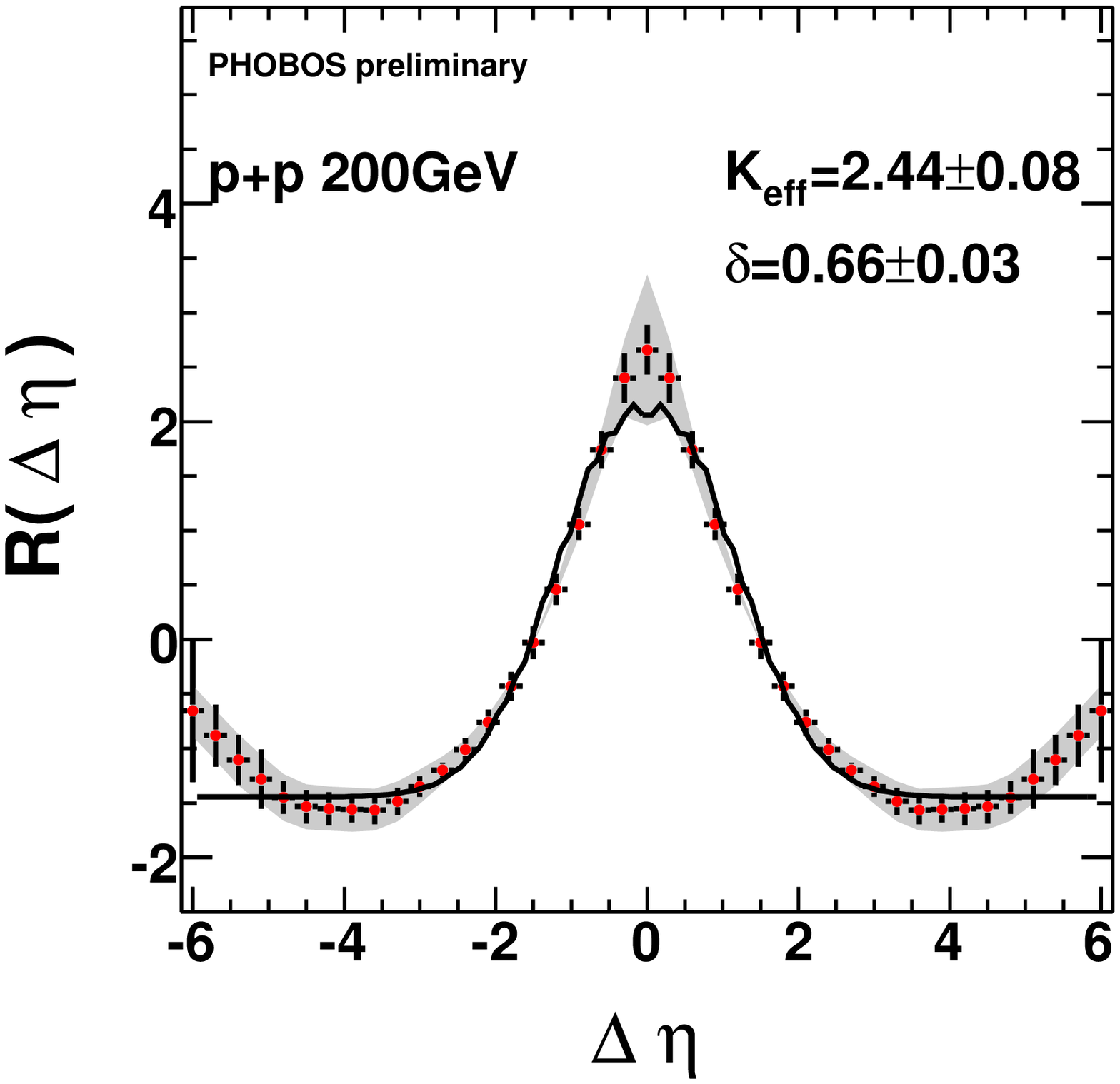}
 }
 \vspace{-0.3cm}
 \caption{1D two-particle rapidity correlation function in 
          ($\Delta \eta$) for p+p collisions at $\sqrt{s}$ = 200~GeV 
          together with a fit from a cluster model. The fit 
          parameters are: $K_{\rm eff}$=2.44$\pm$0.08 and 
          $\delta$=0.66$\pm$0.03 with 90\% C.L. errors}
 \label{pp200_eta_corrected_fit}
\end{minipage}
\end{figure}

Fig.~\ref{pp200_2D_corrected} shows the two-particle inclusive correlation function in p+p 
collisions at $\sqrt{s}$ = 200~GeV as a function of $\Delta \eta$ and $\Delta \phi$. 
To achieve the correlations between primary particles, a set of correction procedures has 
been applied, based on MC simulations. The complex correlation structure suggests that the 
short range correlation is approximately Gaussian in $\Delta \eta$ and persists over the full 
$\Delta \phi$ range, becoming broader toward higher $\Delta \phi$. If clusters are the precursors 
to the final measured hadrons, a high $p_{T}$ cluster would contribute to a narrow peak at 
the near-side ($\Delta \phi$ near $0^{o}$) region of the correlation function in Fig.~\ref{pp200_2D_corrected}, 
while a lower $p_{T}$ cluster will contribute to the broader away-side.

To study these features quantitatively, the two-dimensional (2D) correlation function is 
projected into a one-dimensional (1D) correlation function $R(\Delta \eta)$ shown in 
Fig.~\ref{2pcorr_clusterfitting_incl}. It takes a form derived in Ref. \cite{cluster_fit} 
in an independent cluster emission model:

\begin{equation}
\label{2pcorr_clusterfitting_incl}
R(\Delta \eta)=\alpha\left[\frac{\Gamma(\Delta \eta)}{B(\Delta \eta)}-1\right]   
\end{equation}

\noindent where $B(\Delta \eta)$ is the background distribution obtained by event-mixing. 
Here the multiplicity dependence of $B(\Delta \eta)$ is not considered since the multiplicity 
distribution in p+p collisions is relatively narrow with $\sigma(n)/<n>$ of around 0.25. 
The parameter $\alpha=\frac{<K(K-1)>}{<K>}$ contains the information about the cluster 
size $K$ and $\Gamma(\Delta \eta)$ is a Gaussian function $\propto exp{(- (\Delta \eta)^{2}/(4\delta^{2}))}$ 
characterizing the correlation of particles produced by a single cluster, where $\sqrt{2}\delta$ 
corresponds to the decay width of the clusters. The effective cluster multiplicity, 
or ``size'' is defined to be $K_{\rm eff}=\frac{<K(K-1)>}{<K>}+1=<K>+\frac{\sigma_{K}^{2}}{<K>}$.
Of course, without any knowledge of the distribution of $K$, it is impossible to directly 
measure the average cluster size $<K>$. However, by a $\chi^{2}$ fit of Eq.~\ref{2pcorr_clusterfitting_incl} 
to the measured two-particle correlation function, an example of which shown in Fig.~\ref{pp200_eta_corrected_fit},
the effective cluster size $K_{\rm eff}$ can be estimated, as well as the cluster decay width $\sqrt{2}\delta$. 
A $K_{\rm eff}$ of about 2.5 indicates that on average every charged particle is strongly 
correlated with another 1.5 particles, if it is assumed that $\sigma_{K}^{2}$=0. 

\begin{figure}[t]
\captionsetup[figure]{margin=0.1cm,font=small}
 \vspace{-1.0cm}
\begin{minipage}[t]{0.40\textwidth}
 \centerline{
  \includegraphics[width=\textwidth]{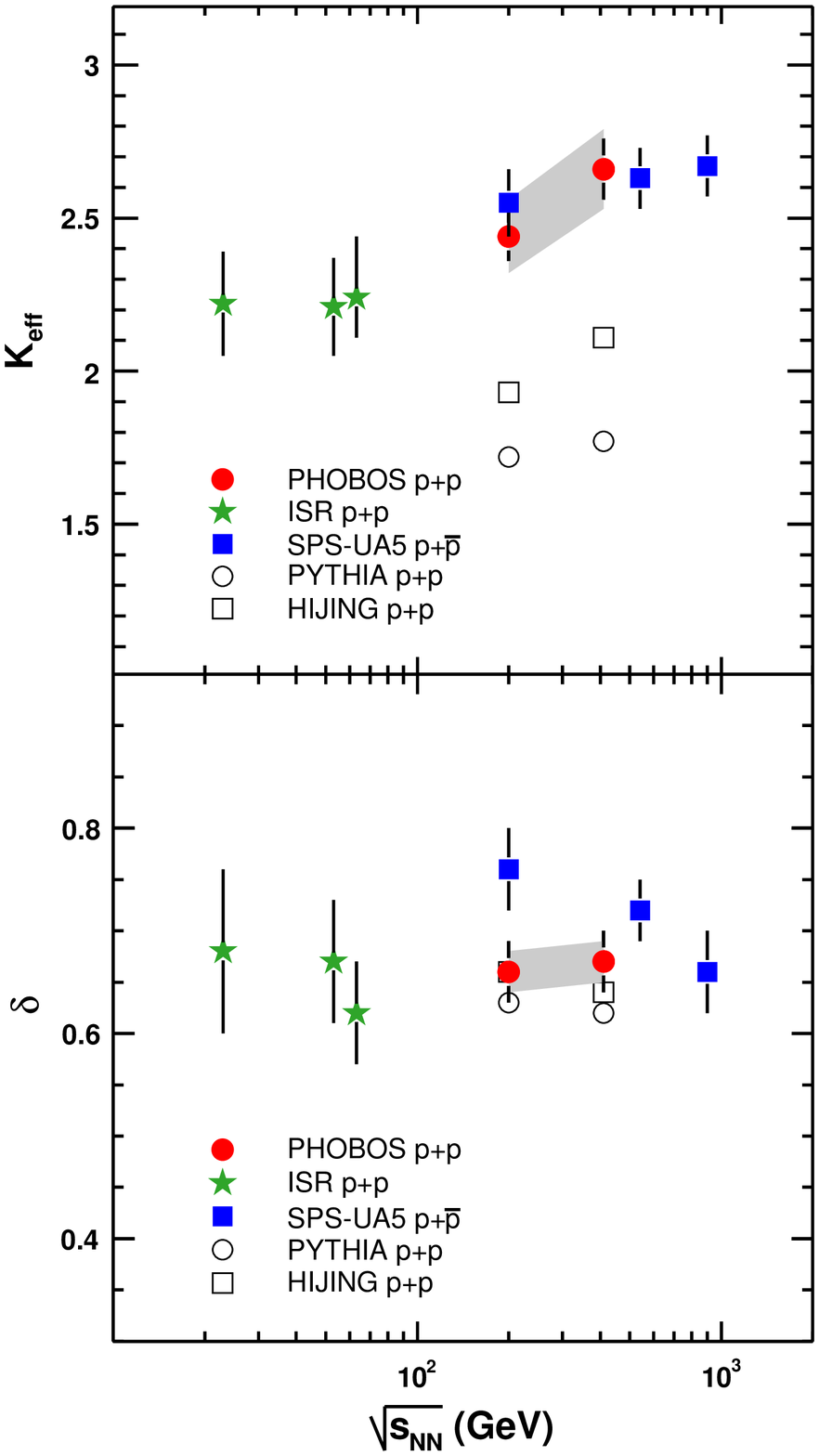}
 }
 \vspace{-0.5cm}
 \caption{$K_{\rm eff}$ (top) and $\delta$ (bottom) as a function of $\sqrt{s}$ 
          measured by PHOBOS (solid circle), UA5 \cite{UA5_3energy} 
          (solid square) and ISR \cite{ISR_twolowenergy, ISR_63GeV} 
          (solid star) experiments in p+p and p+\={p} collisions.}
 \label{cluster_sqrts}
\end{minipage}
\hspace{\fill}
%
\begin{minipage}[t]{0.48\textwidth}
 \centerline{
  \includegraphics[width=\textwidth]{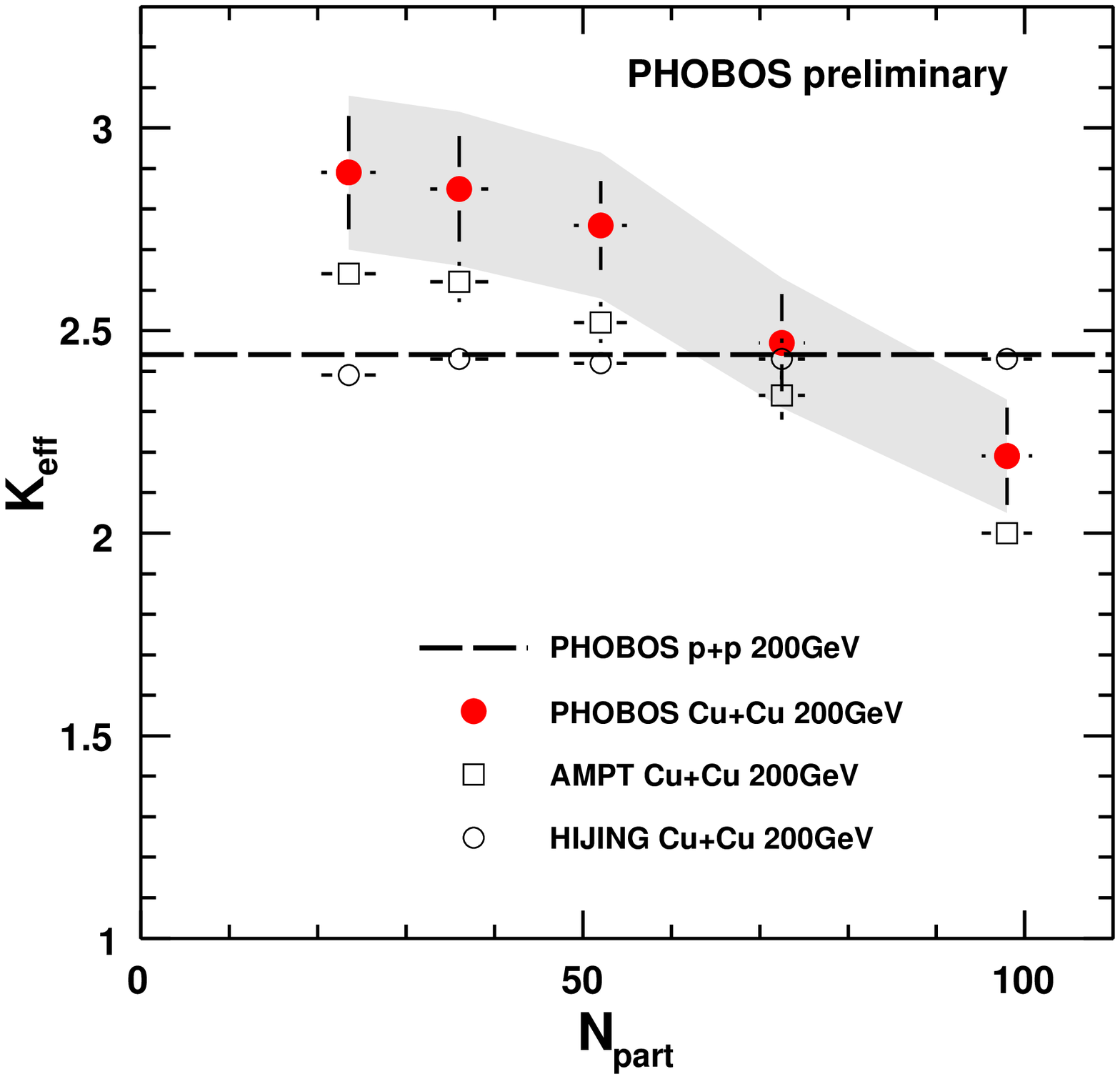}
 }
 \vspace{-0.5cm}
 \caption{ Effective cluster size $K_{\rm eff}$ as a function 
           of $N_{part}$ in Cu+Cu collisions at $\sqrt{s_{_{NN}}}$ = 
           200~GeV (solid circle).The dashed line shows the corresponding 
           value in p+p collisions measured by PHOBOS.}
 \label{k_npart_CuCu_ppline}
\end{minipage}
\end{figure}

Fig.~\ref{cluster_sqrts} shows the collision energy dependence of $K_{\rm eff}$ and $\delta$.  
PHOBOS data at $\sqrt{s}$ = 200~GeV and 410~GeV are consistent with the previous 
UA5 measurements \cite{UA5_3energy} and show a significant increase of the cluster size 
with energy. The cluster decay width is essentially constant with collision energy. 
The observed cluster size exceeds the expectation from resonance decay (about 1.7) \cite{UA5_3energy} 
even at very low collision energy \cite{ISR_twolowenergy, ISR_63GeV}, suggesting additional 
sources of short-range correlations are required to account for the experimental results.

 \vspace{-0.5cm}
\section{Two-particle correlations in Cu+Cu collisions}
In Cu+Cu collisions, a $\cos(2\Delta \phi)$ dependence is observed, and attributed to 
collective flow. However, in this work, the $\Delta \phi$ dependence is integrated out 
in order to just study the cluster properties in pseudorapidity. As was explained above, 
the two-particle pseudorapidity correlation function in Cu+Cu is fit by 
Eq.~\ref{2pcorr_clusterfitting_incl}. The resulting effective cluster size as a function 
of participating nucleons ($N_{part}$) is shown in Fig.~\ref{k_npart_CuCu_ppline}
for Cu+Cu collisions at $\sqrt{s_{_{NN}}}$ = 200~GeV. The dashed line indicates the 
value found in $\sqrt{s}$ = 200~GeV p+p collisions, and suggests that the cluster properties 
are similar in the two systems. This seems reasonable if clusters are produced in the 
late stages of the collision evolution and mainly reflect the hadronization process.
However, it is observed that the cluster size decreases with increasing collision centrality.
In comparing the data with dynamical models, it is found that AMPT gives the same 
qualitative trend as the data, but is systematically lower in magnitude.  By contrast, 
HIJING remains constant with increasing centrality.  Further comparison of Cu+Cu and 
Au+Au data with p+p should provide more information on the dynamical origins of the centrality 
dependence seen in the Cu+Cu data.

\section{Conclusion}

In conclusion, the two-particle correlation function for inclusive charged particles 
has been studied over broad range in $\eta$ and $\phi$. In particular, it is shown that 
the observed short-range correlations in pseudorapidity have a natural interpretation 
in terms of clusters. In this approach, multiple particles are understood to be emitted 
close together in phase space, with a typical cluster size of 2-3 in p+p collisions.
In Cu+Cu, clusters have a similar size but show a non-trivial decrease in size with 
increasing centrality.  Future work will extend this study by providing a comprehensive 
study of two-particle correlations in p+p, d+Au, Cu+Cu and Au+Au reactions.

\section*{Acknowledgments}

This work was partially supported by U.S. DOE grants 
DE-AC02-98CH10886,
DE-FG02-93ER40802, 
DE-FG02-94ER40818,  
DE-FG02-94ER40865, 
DE-FG02-99ER41099, and
DE-AC02-06CH11357, by U.S. 
NSF grants 9603486, 
0072204,            
and 0245011,        
by Polish KBN grant 1-P03B-062-27(2004-2007),
by NSC of Taiwan Contract NSC 89-2112-M-008-024, and
by Hungarian OTKA grant (F 049823).

\end{document}